\documentstyle[twocolumn,seceq,letter,epsf]{jpsj}
\def\runtitle{Impurity Site NMR Relaxation in Unconventional Superconductors}
\def\runauthor{Masashige {\sc Matsumoto}}

\title
{
Impurity Site NMR Relaxation in Unconventional Superconductors
}

\author
{
Masashige {\sc Matsumoto}
}

\inst
{
Department of Physics, Faculty of Science, Shizuoka University, 836 Oya,
Shizuoka 422-8529, Japan
}

\recdate
{May 8, 2001}

\abst
{
Impurity nuclear spin relaxation is studied theoretically.
A single impurity generates a bound state localized around the impurity atom
in unconventional superconductors.
With increasing impurity potential,
the relaxation rate $T_1^{-1}$ is reduced by the impurity potential.
However, it has a peak at low temperatures due to the impurity bound state.
The peak disappears at non-impurity sites.
The impurity site NMR measurement detecting a local electronic structure just on the impurity atom
is very useful for identifying  the unconventional pairing states.
}

\kword
{
NMR, NQR, unconventional superconductivity, impurity
}
\begin{document}
\sloppy
\maketitle
%%%%%%%%%%%%%%%%%%%%%%%%%%%%%%%%%%%%%%%%%%%%%%%%%%%%%%%%%%%%%%%%%%%%%%%%%%%%%%%%
\renewcommand{\theequation}{\arabic{equation}}
%%%%%%%%%%%%%%%%%%%%%%%%%%%%%%%%%%%%%%%%%%%%%%%%%%%%%%%%%%%%%%%%%%%%%%%%%%%%%%%%

%%%%%%%%%%%%%%%%%%%%%%%%%%   Introduction   %%%%%%%%%%%%%%%%%%%%%%%%%%%%%%%%%%%%
The study of unconventional superconductors
has become one of the most attractive issues in recent condensed matter physics.
These materials include heavy fermion compounds, high-$T_c$ cuprates,
organic superconductors and the recently discovered Sr$_2$RuO$_4$.
A great effort has been made to detect the unconventional superconductivity by several probes:
specific heat measurement, tunneling spectroscopy, nuclear magnetic resonance (NMR),
$\pi$-junction and so on.
NMR experiments were important in the early stage
of studying the unconventional superconductivity.
It is well known that $T_1^{-1}$ in $s$-wave superconductors
has a Hebel-Slichter peak just below $T_c$,
while it does not exhibit this feature in unconventional superconductors.
On the other hand, it was also reported that the peak can be suppressed
by the strong damping effect even for a $s$-wave state.
\cite{Kitaoka}

To clarify this point,
Ishida et al. performed a Cu site (bulk site) NMR experiment
with Zn-doped YBa$_2$Cu$_3$O$_{7-\delta}$ (YBCO).
\cite{Ishida93}
They found that $T_1^{-1}$ is proportional to $T$ at low temperatures.
Non-magnetic impurities do not break the $s$-wave superconductivity.
\cite{Anderson}
However, they break unconventional superconductivity and yield low-energy impurity states.
\cite{Hirschfeld,Schmitt-Rink,Hotta}
The observed $T$ linear dependence of $T_1^{-1}$ can therefore
be understood as evidence of extended low-energy impurity states
and supports an unconventional pairing state for YBCO.
In this case, the effect of many impurities was important.

A single impurity can also break the unconventional superconductivity locally
and generate a bound state around it.
\cite{Matsumoto1,Balatsky,Onishi}
Recently, Pan et al. observed such a low-energy state around a Zn atom
in Bi$_2$Sr$_2$CaCu$_2$O$_{8+\delta}$
using a low-temperature scanning tunneling microscope (STM),
showing a fourfold symmetrical structure
which is consistent with a $d_{x^2-y^2}$-wave pairing state.
\cite{Pan,Yazdani}
Nishida et al. also observed the fourfold structure
around a columnar defect by a STM.
\cite{Nishida}
At present, STM is the only instrument
which can reveal the local electronic structure around a single impurity atom.

Usually, NMR is used to probe bulk electronic structure.
For example, Cu site NMR measurements for high-$T_c$ cuprates
detect local electronic structure at the Cu atom.
Recently, impurity site NMR experiments were performed by several groups.
\cite{Ishida96,Bobroff}
However, there is, so far, no theory for the impurity site NMR.
The purposes of this paper are (1) to address the theory for the impurity site NMR
and (2) to propose a new NMR experiment
which probes the local quasiparticle states around a single impurity atom.
In this study we concentrate our attentions on a single impurity problem,
since we can solve it exactly.
For unconventional superconductivity,
we show that the impurity site NMR $T_1^{-1}$ has a peak,
while the peak disappears at the bulk site (non-impurity site).
This peak at the impurity site is a common feature of all types of unconventional superconductors.
The impurity site NMR can reveal the local quasiparticle structure as well,
and provides us with a new experimental method
for identifying unconventional superconductivity.
Recently, NMR $T_1^{-1}$ in a vortex state was studied theoretically.
\cite{Takigawa,Morr}
They proposed that the NMR can be used as a site-selective probe
by controlling the resonance frequency.
Thus, NMR is gradually attracting attention as a local probe.
%with the development in experiment.

%%%%%%%%%%%%%%%%%%%%%%%%%%   Formulation   %%%%%%%%%%%%%%%%%%%%%%%%%%%%%%%%%%%%%
In the unconventional pairing case,
the superconducting order parameter is suppressed around the impurity.
However, we present our theory assuming a uniform order parameter,
since we can capture the essential physics clearly without any complex analysis.
Actually, we have performed numerical self-consistent calculations
to include the spatial dependence of the order parameter.
We have confirmed that the NMR $T_1^{-1}$ has no qualitative difference
between in the uniform and non-uniform order parameter cases.
The most important point in the impurity site NMR is the existence of the impurity bound state,
which does not depend on details of unconventional pairing states.

For the unconventional pairings,
we focus on $p_x\pm\ {\rm i} p_y$-wave and $d_{x^2-y^2}$-wave pairing states.
The $p_x\pm\ {\rm i} p_y$-wave is the simplest and most essential symmetry
suggested for the Sr$_2$RuO$_4$ superconductor.
\cite{Maeno,Rice,Baskaran}
This state is a spin triplet state and breaks the time-reversal symmetry.
\cite{Luke,Ishida98}
For simplicity, we assume that the superconductor is basically
two-dimensional and has a cylindrical Fermi surface.
The Matsubara Green function in a $2\times 2$ matrix form is given by
\cite{Matsumoto1}
\begin{eqnarray}
&&{\hat G}(  {\rm i} \omega_m,\mbox{\boldmath$r$},\mbox{\boldmath$r$}')
= {\hat G}_0({\rm i} \omega_m,\mbox{\boldmath$r$},\mbox{\boldmath$r$}')
                       + {\hat G}_0({\rm i} \omega_m,\mbox{\boldmath$r$},0) U_0 {\hat \tau}_3 \cr
&&~~~~~~~~~~~\times{1 \over 1 - {\hat G}_0({\rm i} \omega_m,0,0) U_0 {\hat \tau}_3}
{\hat G}_0({\rm i} \omega_m,0,\mbox{\boldmath$r$}'),
\end{eqnarray}
where a single impurity is located at the origin of the coordinate.
$U_0$ represents the strength of the short-range impurity potential.
$\omega_m$ is the fermion Matsubara frequency
and ${\hat {\tau_i}}$ $(i=1,2,3)$ represents the Pauli matrix in a charge space.
For simplicity we use $\hbar=1$ and $k_{\rm B}=1$ units throughout this letter.
${\hat G}_0$ is the non-perturbed ($U_0=0$) Green function given by
\begin{equation}
{\hat G}_0({\rm i} \omega_m,\mbox{\boldmath$r$},\mbox{\boldmath$r$}')
= -{ 1 \over \Omega} \sum_{\mbox{\boldmath$k$}}
   {\rm e}^{{\rm i} \mbox{\footnotesize \boldmath $k$}
    \cdot (\mbox{\boldmath$r$}-\mbox{\boldmath$r$}')}
     {{\rm i} \omega_m + \epsilon_{\mbox{\footnotesize \boldmath $k$}}{\hat \tau}_3
   + {\hat \Delta}_{\mbox{\footnotesize \boldmath $k$}}
   \over
     \omega_m^2 + \epsilon_{\mbox{\footnotesize \boldmath $k$}}^2
   + \Delta_{\mbox{\footnotesize \boldmath $k$}}^2}.
\end{equation}
Here, ${\hat \Delta}_{\mbox{\footnotesize \boldmath $k$}}$
is equal to $\Delta_{\mbox{\footnotesize \boldmath $k$}}{\hat \tau}_1$
and expresses the momentum-dependent order parameter.
For the $p_x\pm {\rm i} p_y$-wave ($d_{x^2-y^2}$-wave),
$\Delta_{\mbox{\footnotesize \boldmath $k$}}=\Delta_p(T){\rm e}^{{\rm i} \theta_k}$
[$\Delta_{\mbox{\footnotesize \boldmath $k$}}=\Delta_d(T)\cos(2\theta_k)$].
Here, $\theta_k$ is the angle of the Fermi wave vector measured from the $k_x$ axis.
We solve the gap equation in the bulk region
and use the temperature-dependent order parameter $\Delta_p(T)$ ($\Delta_d(T)$)
which is real and positive for the $p_x\pm {\rm i} p_y$-wave ($d_{x^2-y^2}$-wave).
$\epsilon_{\mbox{\footnotesize \boldmath $k$}}$ is the kinetic energy
and $\Omega$ represents the volume of the system.
NMR relaxation occurs via the hyperfine interaction
between the nuclear spin and the conduction electrons.
NMR $T_1^{-1}$ at the impurity site is given in an explicit form \cite{Leadon}
\begin{eqnarray}
&&T_1^{-1}=2\pi(\frac{4\pi}{3})^2(\gamma_e\gamma_n)^2 W, \cr
&&W=\int {\rm d} E { a_{11}(E) a_{22}(-E)-a_{12}(E) a_{21}(-E) \over 1+\cosh(E/T)},
\label{eqn:T1}
\end{eqnarray}
where $W$ is proportional to the nuclear spin flip transition probability.
$\gamma_e$ and $\gamma_n$ are the gyromagnetic ratios
for the electron and nucleon, respectively.
$a_{ij}(E) = -{\rm Im}\bigl[G_{ij}({\rm i} \omega_m \rightarrow E
             + {\rm i} \delta,0,0)\bigr]/\pi$,
where the subscript of $a_{ij}$ represents the matrix element of the Green function.
Here, we have omitted the $u_0$ dependence in $a_{ij}$ for simplicity,
where $u_0=\pi N_0 U_0$ ($N_0$ is the density of states per volume at the Fermi energy).
$\delta$ is a positive, small number which expresses a finite level broadening.

\begin{figure}[t]
\begin{center}
\begin{minipage}{6.2cm}
\epsfxsize=6.2cm
\epsfbox{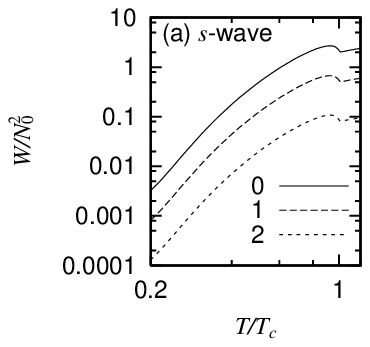}
\end{minipage}
\begin{minipage}{6.2cm}
\epsfxsize=6.2cm
\epsfbox{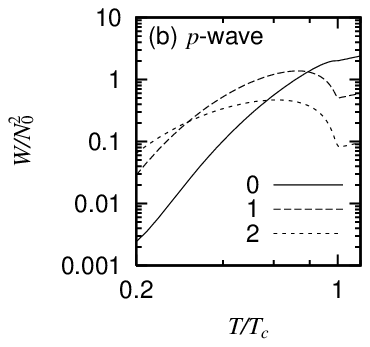}
\end{minipage}
\end{center}
\caption{
Temperature dependence of $W$ with various $u_0$ with a fixed damping rate $\delta/T_c=0.1$.
The number represents a value of $u_0$.
(a) and (b) are for the $s$-wave and $p_x \pm {\rm i} p_y$-wave, respectively.
The peak of $T_1^{-1}$ in (b) becomes higher with the decrease of $\delta/T_c$
[see eq. (\ref{eqn:Wb})].
The value of $\delta/T_c$ does not change the temperature dependence of $T_1^{-1}$.
}
\label{fig:1}
\end{figure}

%%%%%%%%%%%%%%%%%%%%%%%%%%   s-wave   %%%%%%%%%%%%%%%%%%%%%%%%%%%%%%%%%%%%%%%%%%
In the $s$-wave case,
NMR $T_1^{-1}$ has the Hebel-Slichter peak just below $T_c$ at the non-impurity site ($u_0=0$).
At the impurity site ($u_0\neq 0$),
$a_{ij}$ scales as $a_{ij}(E,u_0)=a_{ij}(E,0)/(1+u_0^2)$.
The local density of states at the impurity is then simply reduced,
keeping the same energy dependence.
This is consistent with Anderson's theorem.
\cite{Anderson}
In the same manner, $W$ scales as
\begin{equation}
W(u_0,T)=W(0,T)/(1+u_0^2)^2.
\label{eqn:Ws}
\end{equation}
Therefore, $T_1^{-1}$ is simply reduced at the impurity site,
while the temperature dependence does not change
between the impurity site and the bulk site for the $s$-wave, as shown in Fig. \ref{fig:1}(a).
%%%%%%%%%%%%%%%%%%%%%%%%%%   Fig. 1   %%%%%%%%%%%%%%%%%%%%%%%%%%%%%%%%%%%%%%%%%%
%%%%%%%%%%%%%%%%%%%%%%%%%%%%%%%%%%%%%%%%%%%%%%%%%%%%%%%%%%%%%%%%%%%%%%%%%%%%%%%%

\begin{figure}[t]
\begin{center}
\begin{minipage}{6.2cm}
\epsfxsize=6.2cm
\epsfbox{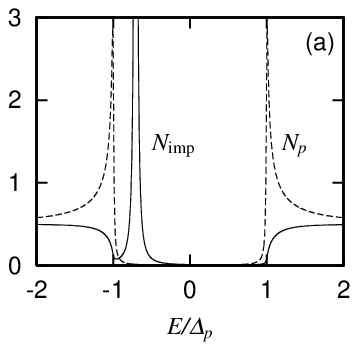}
\end{minipage}
\begin{minipage}{6.2cm}
\epsfxsize=6.2cm
\epsfbox{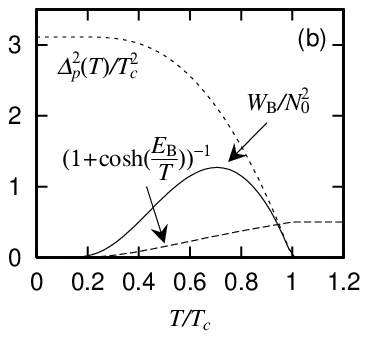}
\end{minipage}
\end{center}
\caption{
(a) Dimensionless local density of states for the $p_x\pm i p_y$-wave state.
$N_p$ is multiplied by $(1+u_0^2)^{-1}=N_{\rm imp}(\infty)$ for convenience.
Parameters are chosen as $u_0=1$ and $\delta/\Delta_p=0.01$.
(b) Temperature dependence of $W_{\rm B}$.
Parameters are chosen as $u_0=1$ and $\delta/T_c=0.1$.
}
\label{fig:2}
\end{figure}

%%%%%%%%%%%%%%%%%%%%%%%%%%   p-wave   %%%%%%%%%%%%%%%%%%%%%%%%%%%%%%%%%%%%%%%%%%
For the $p_x \pm {\rm i} p_y$-wave,
we show the results in Fig. \ref{fig:1}(b).
In this case,
the temperature dependence at the impurity site
is quite different from that of the bulk site.
In the unconventional pairing case, both $a_{12}$ and $a_{21}$ are zero,
since pair electrons cannot possess the same position.
Therefore the coherence factor vanishes for unconventional pairings,
resulting in the absence of the Hebel-Slichter peak.
$W$ is then expressed by the following simple form:
\begin{equation}
W=N_0^2\int {\rm d}E \frac{N_{\rm imp}^2(E)}{1+\cosh(E/T)},
\label{eqn:W}
\end{equation}
where $N_{\rm imp}(E)=a_{11}(E)/N_0=a_{22}(-E)/N_0$
is the dimensionless local density of states at the impurity atom.
For the $p_x\pm {\rm i} p_y$-wave, a bound state appears around the single impurity.
The energy position of the bound state is given by
$E_{\rm B}=-{\rm sgn}(u_0) \Delta_p(T) /\sqrt{1+u_0^2}$.
\cite{Okuno}
In fact, an impurity effect is observed in Sr$_2$RuO$_4$
as a reduction of its $T_c$.
\cite{Mackenzie}
$N_{\rm imp}$ in eq. (\ref{eqn:W}) for the $p_x\pm {\rm i} p_y$-wave
has the following analytic form:
\begin{eqnarray}
N_{\rm imp}(E)
&=&\frac{N_p(E)}{1+[u_0 N_p(E)]^2}\theta(|E|-\Delta_p(T)) \cr
&+&\frac{\pi|u_0|\Delta_p(T)}{(1+u_0^2)^{\frac{3}{2}}}\delta(E-E_{\rm B}),
\label{eqn:Np}
\end{eqnarray}
where $N_p(E)=|E|/\sqrt{E^2-\Delta_p^2(T)}$
is the dimensionless $p_x\pm {\rm i} p_y$-wave bulk density of states.
$\theta$ and $\delta$ in eq. (\ref{eqn:Np})
are the Heaviside and $\delta$-functions, respectively.
The first and second terms in eq. (\ref{eqn:Np})
are the continuum and bound states, respectively.
Introducing $u_0$, we can see in Fig. \ref{fig:2}(a)
that the density of states from the quasiparticle continuum is decreased,
while that of the impurity bound state increases.
%%%%%%%%%%%%%%%%%%%%%%%%%%   Fig. 2   %%%%%%%%%%%%%%%%%%%%%%%%%%%%%%%%%%%%%%%%%%
%%%%%%%%%%%%%%%%%%%%%%%%%%%%%%%%%%%%%%%%%%%%%%%%%%%%%%%%%%%%%%%%%%%%%%%%%%%%%%%%
As in the $s$-wave case,
$W$ for the $p_x\pm {\rm i} p_y$-wave state is reduced with the increase of $u_0$.
However, as shown in Fig. \ref{fig:1}(b),
it drastically changes at low temperatures due to the impurity bound state.
The bound state contribution to $W$ behaves as
\begin{equation}
W_{\rm B}
\simeq
N_0^2 \Bigl[\frac{\Delta_p(T) u_0}{(1+u_0^2)^{\frac{3}{2}}}\Bigr]^2
\frac{\pi}{2\delta} [1+\cosh(E_{\rm B}/T)]^{-1},
\label{eqn:Wb}
\end{equation}
where $\delta(E-E_{\rm B})=(\delta/\pi)[(E-E_{\rm B})^2+\delta^2]^{-1}$ has been used.
Figure \ref{fig:2}(b) clearly shows that $W_{\rm B}$ has a peak below $T_c$.
Notice that the peak in $W$ in Fig. \ref{fig:1}(b)
is closely related to the local impurity bound state.
It is different from the Hebel-Slichter peak and can be distinguished,
since $T_1^{-1}$ does not exhibit such a peak at the bulk site
for unconventional superconductivity.
The peak appears only at the impurity site.
At low temperatures, impurity nuclear spin relaxation occurs
via the impurity bound state at the impurity site.
As $u_0$ is increased,
the peak position of $W$ shifts to a lower temperature region [see Fig. \ref{fig:1}(b)],
which reflects the energy shift of the bound state.
For a large $u_0$, $W_{\rm B}$ can survive if $\delta$ is sufficiently small.

\begin{figure}[t]
\begin{center}
\begin{minipage}{6.2cm}
\epsfxsize=6.2cm
\epsfbox{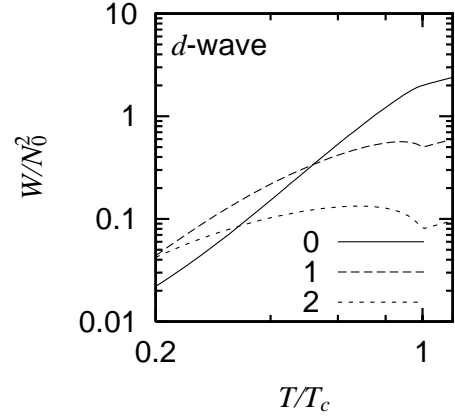}
\end{minipage}
\end{center}
\caption{
Temperature dependence of $W$ with various $u_0$ for the $d_{x^2-y^2}$-wave
with a fixed damping rate $\delta/T_c=0.1$.
The number represents a value of $u_0$.
}
\label{fig:3}
\end{figure}

%%%%%%%%%%%%%%%%%%%%%%%%%%   d-wave   %%%%%%%%%%%%%%%%%%%%%%%%%%%%%%%%%%%%%%%%%%
Next we study the $d_{x^2-y^2}$-wave case,
which is the most favorable pairing symmetry for high-$T_c$ cuprates.
In this case, $N_{\rm imp}$ in eq. (\ref{eqn:W}) takes the following form:
\begin{equation}
N_{\rm imp}(E)=\frac{1}{u_0^2}\frac{N_d(E)}{[A(E)+1/u_0]^2+N_d^2(E)},
\end{equation}
where $A(E)$ and $N_d(E)$ are the real and imaginary parts of
$\bigl\langle{(E+ {\rm i} \delta) / \sqrt{(-{\rm i} E+\delta)^2
+|\Delta_{\mbox{\footnotesize \boldmath $k$}}|^2}}\bigr\rangle_{\rm FS}$,
respectively.
Here, $\langle \cdots \rangle_{\rm FS}$ represents an average over the Fermi surface.
The energy position of the bound state is determined by the zero of $A(E)+1/u_0$
and its effective broadening is $N_d(E)$
which is the dimensionless $d_{x^2-y^2}$-wave bulk density of states.
Similar to the $p_x \pm {\rm i} p_y$-wave case,
the impurity site ($u_0 \neq 0$) NMR $T_1^{-1}$ for the $d_{x^2-y^2}$-wave has a peak
due to the impurity bound state, as shown in Fig. \ref{fig:3}.
%%%%%%%%%%%%%%%%%%%%%%%%%%   Fig. 3   %%%%%%%%%%%%%%%%%%%%%%%%%%%%%%%%%%%%%%%%%%
%%%%%%%%%%%%%%%%%%%%%%%%%%%%%%%%%%%%%%%%%%%%%%%%%%%%%%%%%%%%%%%%%%%%%%%%%%%%%%%%

%%%%%%%%%%%%%%%%%%%%%%%%%%   Discussion   %%%%%%%%%%%%%%%%%%%%%%%%%%%%%%%%%%%%%%
The important point in this letter is that
a single impurity generates a bound state in unconventional superconductors,
and that the NMR experiment can detect the local impurity bound state
by the impurity site $T_1^{-1}$ measurement.
We emphasize that the peak in $T_1^{-1}$ at low temperatures
in the impurity site NMR is a common feature
exhibited by all types of unconventional superconductors.
It does not depend on the details of unconventional pairing states.
We have demonstrated this
by examining the $p_x \pm {\rm i} p_y$-wave and $d_{x^2-y^2}$-wave pairing states.
The former is a fully gapped unconventional state,
while the latter is a gapless one.

Recently Ishida et al. carried out NMR and NQR measurements on CeRu$_2$
with 1\% Al (impurity) substitution for Ru.
\cite{Ishida96}
They reported that the $T_1^{-1}$ at the Al site
was reduced to 10\% of that of the Ru (bulk) site.
The strength of the impurity scattering can be estimated as $|u_0|=1.5$
by using eq. (\ref{eqn:Ws}).
If the pairing is unconventional,
we can expect the appearance of a peak below $T_c$ for $|u_0|=1.5$,
as shown in Figs. \ref{fig:1}(b) and \ref{fig:3}.
However, the Al site $T_1^{-1}$ does not exhibit such a peak.
It is simply reduced, keeping the same exponential temperature dependence as that of the Ru site.
This feature is consistent with our theory of impurity site NMR for the $s$-wave.

Very recently, Bobroff et al. succeeded in Li (impurity) site NMR measurement
in YBCO with Li substituted for Cu.
\cite{Bobroff}
The Li concentrations of the samples were 0.85\% and 1.86\% per CuO$_2$ layer.
For YBCO, Li induces a magnetic moment around it.
Unfortunately, it is difficult to extract the contribution from the conduction electrons in $T_1$,
since the induced moment also generates the relaxation.
However, this Li site experiment shows that
the recent NMR experiment has enough sensitivity
to detect $T_1$ at the impurity site in unconventional superconductors.
The impurity concentration is about 1\% for both CeRu$_2$ and YBCO.
These experiments indicate
that the impurity site NMR is possible at least for such impurity doping.

For the bulk site NMR $T_1$ measurement,
it is important to observe whether the Hebel-Slichter peak is present or not.
Examining the temperature dependence (exponential or power) at low temperatures is also important.
(I) In the absence of the Hebel-Slichter peak,
it is difficult to distinguish unconventional pairings from the $s$-wave
for the bulk site measurement,
since strong damping can suppress the coherence peak for the $s$-wave.
(II) The temperature dependence of the bulk site $T_1^{-1}$ is not useful
for distinguishing fully gapped unconventional states from the $s$-wave.
We cannot distinguish gapless unconventional states from the anisotropic $s$-wave, either.
In contrast to the bulk site measurement,
the impurity site measurement can distinguish for both (I) and (II) cases.
The feature of unconventional pairings appears as a peak in $T_1^{-1}$ only at the impurity site,
while the peak disappears at the bulk site.
On the other hand, for the $s$-wave,
the temperature dependence of $T_1^{-1}$ does not change
for either impurity or bulk sites [see eq. (\ref{eqn:Ws}) and Fig. \ref{fig:1}(a)].
If we combine the two (bulk and impurity sites) NMR measurements,
we can identify the unconventional pairing without ambiguity.
Thus, the impurity site NMR is a powerful experimental method
for probing unconventional superconductivity.

Our theory for a single impurity is applicable to a case with low impurity concentration.
With the increase of the impurity concentration,
extended impurity states are formed by the local impurity bound states,
resulting in additional continuum energy levels below the energy gap.
In this case,
the impurity effect appears as a constant $T_1 T$ at low temperatures
for both the impurity and bulk sites.
However, the impurity site measurement is expected to exhibit the impurity effect
more obviously than the bulk site,
since the electronic states at the impurity site are strongly affected by the impurity atom.
We note that the peak in $T_1^{-1}$ of our theory
for the single impurity (low impurity concentration)
is much more drastic than the constant $T_1 T$,
both of which are evidence of unconventional superconductivity.

The local electronic structure around the impurity atom
reflects the superconducting pairing symmetry.
So far, STM is the only probe of the local impurity state.
However, STM experiments are strongly affected by the surface roughness.
We note that the impurity site NMR can be used as a local probe
regardless of annoying surface conditions.

In conclusion, we have addressed the theory for the impurity site NMR $T_1^{-1}$
in unconventional superconductors.
We have concentrated our attention on a single impurity effect.
There are two characteristic points of the impurity site NMR.
The first one is that the relaxation rate $T_1^{-1}$ is reduced by the impurity potential.
The second is the peak in $T_1^{-1}$.
It appears as direct result of the relaxation process via the impurity bound state.
On the other hand, the peak vanishes at a non-impurity site.
This is the remarkable difference in $T_1^{-1}$
between the impurity site and non-impurity site NMR in unconventional superconductors.
In contrast,
the temperature dependence of $T_1^{-1}$ does not change
between the two sites for the conventional $s$-wave.
Therefore, observation of the peak by the impurity site NMR measurement
can provide strong evidence of an unconventional superconducting state.

%%%%%%%%%%%%%%%%%%%%%%%%%%%%%%%%%%%%%%%%%%%%%%%%%%%%%%%%%%%%%%%%%%%%%%%%%%%%%%%%
The author expresses his sincere thanks to Y. Okuno and M. Sigrist
for many discussions on the impurity effect in unconventional superconductors.
He also thanks N. Hayashi and M. Takigawa
for interesting discussions on NMR $T_1^{-1}$.
He is grateful for the helpful discussions with H. Alloul and K. Ishida regarding NMR experiments.
He would like to thank M. Koga for his helpful comments
and critical reading of the manuscript.
This work was supported by JSPS for Encouragement of Young Scientists (No. 10740169).
%%%%%%%%%%%%%%%%%%%%%%%%%%%%%%%%%%%%%%%%%%%%%%%%%%%%%%%%%%%%%%%%%%%%%%%%%%%%%%%%

%%%%%%%%%%%%%%%%%%%%%%%%%%%%%%%%%%%%%%%%%%%%%%%%%%%%%%%%%%%%%%%%%%%%%%%%%%%%%%%%


\begin{thebibliography}{99}

%%%%%%%%%%%%%%%%%%%%%%%%%%%%%%%%%%%%%%%%%%%%%%%%%%%%%%%%%%%%%%%%%%%%%%%%%%%%%%%%

\bibitem{Kitaoka}
  Y. Kitaoka et al.:
  Physica C {\bf 192} (1992) 272.

\bibitem{Ishida93}
  K. Ishida et al.:
  J. Phys. Soc. Jpn. {\bf 62} (1993) 2803.

\bibitem{Anderson}
  P. W. Anderson:
  J. Phys. Chem. Solids {\bf 11} (1959) 26.

\bibitem{Hirschfeld}
  P. J. Hirschfeld, D. Vollhardt and P. W\"{o}lfle:
  Solid State Commun. {\bf 59} (1986) 111;
  P. J. Hirschfeld, P. W\"{o}lfle and D. Einzel:
  Phys. Rev. B {\bf 37} (1988) 83;
  P. J. Hirschfeld and N. Goldenfeld:
  Phys. Rev. B {\bf 48} (1993) 4219.

\bibitem{Schmitt-Rink}
  S. Schmitt-Rink, K. Miyake and C. M. Varma:
  Phys. Rev. Lett. {\bf 57} (1986) 2575.

\bibitem{Hotta}
  T. Hotta:
  J. Phys. Soc. Jpn. {\bf 62} (1993) 274.

%%%%%%%%%%%%%%%%%%%%%%%%%%%%%%%%%%%%%%%%%%%%%%%%%%%%%%%%%%%%%%%%%%%%%%%%%%%%%%%%

\bibitem{Matsumoto1}
  M. Matsumoto and H. Shiba:
  J. Phys. Soc. Jpn. {\bf 64} (1995) 1703.

\bibitem{Balatsky}
  A. V. Balatsky, M. I. Salkola and A. Rosengren:
  Phys. Rev. B {\bf 51} (1995) 15547;
  M. I. Salkola, A. V. Balatsky and D. J. Scalapino:
  Phys. Rev. Lett. {\bf 77} (1996) 1841.

\bibitem{Onishi}
  Y. Onishi et al.:
  J. Phys. Soc. Jpn. {\bf 65} (1996) 675.

\bibitem{Pan}
  S. H. Pan et al.:
  Nature (London) {\bf 403} (2000) 746.

\bibitem{Yazdani}
  A. Yazdani et al.:
  Phys. Rev. Lett. {\bf 83} (1999) 176.

\bibitem{Nishida}
  N. Nishida et al.:
  Physica B {\bf 284-288} (2000) 967;
  M. Matsumoto, S. Kaneko and N. Nishida:
  J. Phys. Soc. Jpn. {\bf 66} (1998) 105.

\bibitem{Ishida96}
  K. Ishida et al.:
  Z. Naturforsch {\bf 51 a} (1996) 793;
  H. Mukuda et al.:
  J. Phys. Soc. Jpn. {\bf 67} (1998) 2101.

\bibitem{Bobroff}
  J. Bobroff et al.:
  Phys. Rev. Lett. {\bf 86} (2001) 4116.

\bibitem{Takigawa}
  M. Takigawa, M. Ichioka and K. Machida:
  Phys. Rev. Lett. {\bf 83} (1999) 3057;
  M. Takigawa, M. Ichioka and K. Machida:
  J. Phys. Soc. Jpn. {\bf 69} 3943 (2000).

\bibitem{Morr}
  D. K.  Morr and R. Wortis:
  Phys. Rev. B {\bf 61} (2000) R882;
  R. Wortis, A. J. Berlinsky and C. Kallin:
  Phys. Rev. B {\bf 61} (2000) 12342;
  D. K. Morr:
  Phys. Rev. B {\bf 63} (2001) 214509.

%%%%%%%%%%%%%%%%%%%%%%%%%%%%%%%%%%%%%%%%%%%%%%%%%%%%%%%%%%%%%%%%%%%%%%%%%%%%%%%%

\bibitem{Maeno}
  Y. Maeno et al.:
  Nature (London) {\bf 372} (1994) 532.

\bibitem{Rice}
  T. M. Rice and M. Sigrist:
  J. Phys.: Condens. Matter {\bf 7} (1995) L643;
  T. M. Rice:
  Nature (London) {\bf 396} (1998) 627;
  M. Sigrist et al.:
  Physica C {\bf 317-318} (1999) 134.

\bibitem{Baskaran}
  G. Baskaran:
  Physica B {\bf 223-224} (1996) 490.

\bibitem{Luke}
  G. M. Luke et al.:
  Nature (London) {\bf 394} (1998) 558.

\bibitem{Ishida98}
  K. Ishida et al.:
  Nature (London) {\bf 396} (1998) 658.

%%%%%%%%%%%%%%%%%%%%%%%%%%%%%%%%%%%%%%%%%%%%%%%%%%%%%%%%%%%%%%%%%%%%%%%%%%%%%%%%

\bibitem{Leadon}
  R. Leadon and H. Suhl:
  Phys. Rev. {\bf 165} (1968) 596.

\bibitem{Okuno}
  Y. Okuno, M. Matsumoto and M. Sigrist:
  J. Phys. Soc. Jpn. {\bf 68} (1999) 3054.

\bibitem{Mackenzie}
  A. P. Mackenzie et al.:
  Phys. Rev. Lett. {\bf 80} (1998) 161.

%%%%%%%%%%%%%%%%%%%%%%%%%%%%%%%%%%%%%%%%%%%%%%%%%%%%%%%%%%%%%%%%%%%%%%%%%%%%%%%%

\end{thebibliography}
\end{document}